\documentclass[twocolumn,pre,showpacs]{revtex4}

\usepackage[active]{srcltx}

\usepackage{graphics,graphicx,dcolumn,bm,fleqn,epic,eepic,float}
\usepackage{amssymb,amsmath,multirow,rotate,color}

\newcommand{\bA     }{\mbox{\boldmath$A$}}
\newcommand{\bM     }{\mbox{\boldmath$M$}}
\newcommand{\bU     }{\mbox{\boldmath$U$}}

\newcommand{\bX     }{\mbox{\boldmath$X$}}
\newcommand{\bG     }{\mbox{\boldmath$G$}}
\newcommand{\bT     }{\mbox{\boldmath$T$}}
\newcommand{\bJ     }{\mbox{\boldmath$J$}}
\newcommand{\bphi     }{\mbox{\boldmath$\phi$}}
\newcommand{\bpsi     }{\mbox{\boldmath$\psi$}}

\usepackage{color}

\begin{document}
\title{Finite size corrections to the spectrum of regular random  graphs: an \\ analytical solution}
\author{F. L. \surname{Metz}$^{1}$, G. \surname{Parisi}$^{1,2,3}$ and L. \surname{Leuzzi}$^{1,2}$}
\affiliation{$^1$ Dip. Fisica, Universit{\`a} {\em La Sapienza}, Piazzale
  A. Moro 2, I-00185, Rome, Italy \\ 
$^2$ IPCF-CNR, UOS Roma {\em
 Kerberos}, Universit\`a {\em La Sapienza}, Piazzale A. Moro 2, I-00185,
  Rome, Italy \\ 
$^3$ INFN, Piazzale A. Moro 2, 00185, Rome, Italy}
\date{\today}

\begin{abstract}
We develop a thorough analytical study of the $O(1/N)$ correction
to the spectrum of regular random graphs 
with $N \rightarrow \infty$ nodes. The finite size fluctuations of the resolvent are given 
in terms of a weighted series over the contributions coming from loops of all possible
lengths, from which we obtain the isolated 
eigenvalue as well as an analytical expression for the $O(1/N)$ correction to the continuous part
of the spectrum. 
The comparison
between this analytical formula and direct diagonalization
results exhibits an excellent agreement, confirming the correctness of our expression.
\end{abstract}


\maketitle


\section{Introduction} 

Spectral graph theory has established itself as a fundamental tool
to study problems in various disciplines \cite{Cvetbook}.
On the side of physics, the understanding of stationary
and dynamical properties of models defined on random 
graphs depends crucially on the spectral
analysis of the adjacency and the Laplacian matrix of 
the corresponding graph. The average distribution of 
eigenvalues constitutes a primary object of 
interest, due to its wide range of applications. 
Some notable examples include the study of the vibrational spectra of
amorphous solids \cite{giorgio99}, the electronic properties 
of quantum systems \cite{Abouchacra73} and spherical spin models \cite{baxter1982}. 

A central role in spectral graph theory is
played by sparse random regular graphs (RRGs), since they
constitute a benchmark for
analyzing the spectral features of more complex graph structures.
Random regular graphs are constructed
by drawing, from an uniform probability space, simple undirected
graphs where 
all vertices have the same degree.
Sparse RRGs become locally tree-like
when the total number of nodes $N$ grows to infinite, such that
only long loops of length $O(\ln N)$ are present.
Thanks to the absence of degree fluctuations and to the local tree-like 
structure, many spectral properties of RRGs can be analytically studied
using non-rigorous \cite{TimTese,Kabashima2010} as well as rigorous mathematical approaches (see \cite{Dumi2012} and references therein).
In this context, the most prominent example is the average eigenvalue distribution
of the adjacency matrix, which 
converges, for $N \rightarrow \infty$, to a simple analytical
expression known as the Kesten-McKay (KM) law \cite{Kesten,McKay}. 

Much less is known about the finite size fluctuations
of the spectra of sparse random graphs. The
existence of short loops on graphs
with a finite size and the impact of
these topological fluctuations on the spectral properties
is an interesting problem 
on its own right. 
In addition, sparse random graph models
usually lead, due to its local tree-like topology, to a mean-field description of models
defined on finite-dimensional lattices and,
in a certain sense, the construction of a
perturbative expansion in powers of $O(1/N)$ for random
graph models constitutes an indirect route to
study the intricate role of loops 
on their finite-dimensional counterparts. In fact, analogous
ideas have been put forward in the context of
Anderson localization and statistical mechanics of spin systems \cite{Efetov90,Montanari2005,giorgio2006,vincent2007}, where the behavior of 
models defined on finite-dimensional lattices is studied perturbatively
around the mean-field saddle-point corresponding to sparse random graph models.

Here we implement these ideas to  
study the $O(1/N)$ correction to the average eigenvalue
distribution of the adjacency matrix of RRGs, which are simple enough to render
a full analytical study possible.
We show that the $O(1/N)$ correction to the resolvent
of the adjacency matrix 
is given by a sum over loops comprising all
length scales, each loop contributing with a term
proportional to the difference of its effective resolvent
with respect to the resolvent of an infinite closed chain.
Within the replica approach for random matrices \cite{Edwards76,RB88}, this result is derived from an integration of
the $O(1/\sqrt{N})$ fluctuations of a functional order-parameter around its saddle-point
solution, following analogous
steps as those developed recently to the study of finite size
corrections of models with quenched disorder \cite{Carlo2013,Carlo2014}. 
We show how the divergent loop series
can be summed, leading to a compact analytical expression for the 
$O(1/N)$ correction to the KM law. The correctness of this analytical
formula is confirmed by its very good agreement with numerical
diagonalization results. In addition, our approach allows us to identify the largest 
eigenvalue, separated from the continuous band by a gap, as a singularity in the $O(1/N)$
correction to the resolvent.
To our knowledge, a closely related problem has been considered so
far only in some recent works \cite{BenArous2011,Dumitriu2013,Johnson2012}, where 
it is shown rigorously that the fluctuations of the linear eigenvalue functional of RRGs
converge to a random variable defined in terms of a sum over cyclically non-backtracking walks of all
possible lengths.

The rest of the paper is organized as follows. In the
next section we define the ensemble of RRGs. In section
3 we explain how to recast the problem in terms
of a saddle-point integral using the replica method, and
how one can integrate the fluctuations around the
saddle-point solution. In section 4 the loop series for the $O(1/N)$ correction
to the eigenvalue distribution
is obtained in replica symmetry, while the isolated eigenvalue and the
final analytical expression for the finite size correction to the continuous
band, together with a comparison
with direct diagonalization results, are presented in
section 5. In the last section we present some final 
remarks. The appendix A shows more details
on how to derive the saddle-point integral with the replica
method, while the appendix B discusses the correspondence 
between our results and those of reference \cite{Johnson2012}.


\section{The ensemble of random regular graphs} 

Let us consider the adjacency matrix $\bA$ of an undirected random graph containing $N$ nodes or
vertices, without self-loops and multiple edges between adjacent nodes \cite{Cvetbook}. 
The $N \times N$ symmetric random matrix $\bA$ 
specifies the topology of the graph and it is constructed by setting $A_{ij} = 1$ if there is an
edge between nodes $i$ and $j$, and $A_{ij} = 0$ otherwise.
Defining the eigenvalues of $\bA$ as $\lambda_1,\dots,\lambda_N$, the average 
spectral density reads
\begin{equation}
\rho^{(N)} (\lambda) = \left\langle \frac{1}{N} \sum_{\alpha=1}^{N} \delta(\lambda - \lambda_\alpha) \right\rangle
\,,
\end{equation}
with $\langle \dots \rangle$ denoting the ensemble average
over the distribution of $\bA$.
We study an ensemble of random $c$-regular graphs, where each node
is connected to $c \geq 3$ neighbors and the adjacency matrix
is drawn from the distribution
\begin{align}
p(\{ A_{i < j} \}) &= \frac{1}{\mathcal{A}_N} \left[ \prod_{i < j} \left( \frac{c}{N} \delta_{A_{ij},1} + \left(1 -   \frac{c}{N}\right) 
\delta_{A_{ij},0} \right)  \right] \nonumber \\
&\times
\left[\prod_{i=1}^{N} \delta_{c,\sum_{j=1}^{N}A_{ij}}  \right] \,, \qquad A_{ii} = 0\,.
\label{distrc}
\end{align}
The product $\prod_{i < j}$ runs over all distinct pairs of nodes and $\mathcal{A}_N$ is 
the normalization factor. In this model, the probability
that two nodes are connected by an edge is $c/N$, and the
Kronecker $\delta$ ensures that all vertices are adjacent to $c$ neighbors. 

The averaged resolvent associated to $\bA$ can be defined as 
\begin{equation}
R^{(N)}(z) = \frac{1}{N} \left\langle {\rm Tr} \, \bG(z)  \right\rangle \,,
\end{equation}
where the matrix $\bG(z)$ is given by $\bG(z) = (z - \bA)^{-1}$ and 
$z = \lambda - i \eta$ contains the regularizer $\eta>0$. 
The resolvent $R^{(N)}(z)$ is an analytic function in the
lower half sector of the complex plane, except at the points or segments 
of the real axis corresponding to the eigenvalues of $\bA$, at which
$R^{(N)}(z)$ exhibits singularities. In general, the poles of $R^{(N)}(z)$ can be different
than the simple poles exhibited by ${\rm Tr} \, \bG(z)$, since the latter
quantity is the resolvent before the average over the  distribution of $\bA$
is performed. The average distribution of eigenvalues 
is extracted from the limiting procedure 
\begin{equation}  
\rho^{(N)} (\lambda) = \frac{1}{\pi } \lim_{\eta \rightarrow 0^{+}} {\rm Im}\left[ R^{(N)}(z) \right] \,.
\label{Dosfull}
\end{equation}
By introducing the generating function
\begin{align}
Z_N(z) &= \int \left( \prod_{i=1}^N d \phi_i  \right) \exp{\left(-\frac{iz}{2} \sum_{i=1}^{N} \phi_i^2 \right)} \nonumber \\
&\times
\exp{\left( \frac{i}{2} \sum_{ij=1}^{N}  \phi_i A_{ij} \phi_j \right)}\,,
\label{generat}
\end{align}
$R^{(N)}(z)$ is rewritten as follows
\begin{equation}  
R^{(N)}(z) = - \frac{2}{N} \frac{\partial}{\partial z} 
 \left\langle \ln Z_N(z) \right\rangle  \,.
\label{DOS1}
\end{equation}
In this way, we formulate the problem of computing $\rho^{(N)} (\lambda)$
in the language of statistical mechanics of disordered systems. 
According to eqs. (\ref{Dosfull}-\ref{DOS1}), in order to calculate
$\rho^{(N)} (\lambda)$ and its finite size fluctuations,
one needs to study the average energy density of a system
with real valued ``spins''  $\phi_1,\dots,\phi_N$ placed
on the vertices of a random regular graph and interacting
through ferromagnetic couplings. 

With the purpose of computing the average of the ``free-energy''
$\ln Z_N(z)$ over the random graph topology,
we invoke
the replica method \cite{Edwards76,RB88,Dean,K08}
\begin{equation}
R^{(N)}(z) = - 2
 \frac{\partial}{\partial z} \lim_{n \rightarrow 0} \frac{\partial}{\partial n}  
\frac{1}{N} \ln{\left\langle \left[ Z_N(z) \right]^n  \right\rangle}  \,.
\label{rhorepl}
\end{equation}
The idea consists in calculating the average $\langle \dots \rangle$ of
integer powers of the generating function and, once
the limit $N \rightarrow \infty$ is performed, the number of replicas
is analytically continued to $n \rightarrow 0$.
In this setting, the computation of $\left\langle \left[ Z_N(z) \right]^n  \right\rangle$
is written in terms of an integral over an order-parameter
functional which can be solved, in the limit $N \rightarrow \infty$, by 
means of the saddle-point method, leading to the KM distribution. As we will discuss in the
next section, the $O(1/N)$ correction
to $\lim_{N \rightarrow \infty} \rho^{(N)} (\lambda)$ arises from
the fluctuations of the order-parameter around the saddle-point solution.


\section{The saddle-point integral and the fluctuations around the stationary solution}

The average of the replicated generating function is given
by
\begin{eqnarray}
\left\langle \left[ Z_N(z) \right]^n  \right\rangle &=&
\int \left(  \prod_{i=1}^{N} d \bphi_i  \right) \exp{\left[ -\frac{iz}{2} \sum_{i=1}^{N}  \bphi_{i}^2 \right]} \nonumber \\
&\times&
\left\langle  \exp{\left( i \sum_{i < j} A_{ij} \bphi_i . \bphi_j \right) } \right\rangle \,,
\label{Zn1}
\end{eqnarray}
with $\bphi = (\phi^{1},\dots,\phi^{n})$ denoting a vector in
the $n$-dimensional replica space. The average over the distribution
$p(\{ A_{i < j} \})$ is calculated using integral representations
for the Kronecker $\delta$'s in eq. (\ref{distrc}). After expanding the
integrand exponent in eq. (\ref{Zn1}) up to order $O(N^{0})$, site decoupling
is achieved through the introduction of appropriate order-parameters, which
leads to the compact expression (see the appendix \ref{app1}) 
\begin{equation}
\left\langle \left[ Z_N(z) \right]^n  \right\rangle = 
\sqrt{\det{ \left(  c \,   \bU    \right) }} \int \mathcal{D} \Psi \exp{\left(-N S^{(N)} [\Psi]   \right)} \,.
\label{saddint}
\end{equation}
The object $\Psi(\bphi)$ is the functional order-parameter
and $\bU$ can be seen as a matrix in the configuration space
of the replica vectors, with
elements $U(\bphi,\bpsi)  =  \exp{\left( i \bphi.\bpsi \right) } $.
The functional integration measure can be intuitively written 
as  $\mathcal{D} \Psi = \prod_{ \{ \bphi \} } \sqrt{N/2\pi} \, d \Psi(\bphi)$, where
the product runs over all possible values of the
vector $\bphi$. The action $S^{(N)} [\Psi]$ has been
expanded up to order $O(N^{-1})$ 
\begin{equation}
S^{(N)} [\Psi] = S_0 [\Psi] + \frac{1}{N} S_1 [\Psi] \,,
\label{action1}
\end{equation}
where the coefficients are given by
\begin{align}
S_0 [\Psi]& = \frac{c}{2} \int d \bphi \, d \bpsi \, \Psi(\bphi) U(\bphi,\bpsi ) \Psi(\bpsi) - \frac{c}{2} \nonumber \\
&- \ln{\left[ \int d \bphi  H_z( \bphi)    
\Big(  \int d \bpsi \, U(\bphi,\bpsi )  \Psi(\bpsi)    \Big)^c
\right]} \,, \label{S0} \\
S_1 [\Psi]   &= \frac{1}{4}(c^2 + 1) + \frac{(c-1)}{2} \int d \bphi \, r(\bphi) U(\bphi,\bphi )  \nonumber \\
&+ \frac{(c-1)^2}{4} \int d \bphi \, d \bpsi \,
r(\bphi) \left[ U(\bphi,\bpsi ) \right]^2  r(\bpsi)  \nonumber \\
&- \frac{c^2}{2}  \int d \bphi \, d \bpsi \,  \Psi(\bphi) U(\bphi,\bpsi ) \Psi(\bpsi) - \frac{1}{2} \ln{2} .
\end{align}
In the above expressions we have defined
\begin{align}
H_z( \bphi) &= \exp{\left( - \frac{iz}{2}  \bphi^2  \right)} \,, \label{Hdef} \\
r(\bphi) &= \frac{ H_z( \bphi)  \Big(  \int d \bpsi \, U(\bphi,\bpsi )  \Psi(\bpsi)    \Big)^{c-2}}
{  \int d \bphi  H_z( \bphi)  \Big(  \int d \bpsi \, U(\bphi,\bpsi )  \Psi(\bpsi)    \Big)^{c}    } \,.
\label{eqr}
\end{align}
The details involved in the derivation of eqs. (\ref{saddint}-\ref{eqr}) are discussed
in the appendix \ref{app1}.

In the limit $N \rightarrow \infty$, the integral in eq. (\ref{saddint}) is dominated
by the stationary solution $\Psi_s(\bphi)$ fulfilling
\begin{equation}
\frac{\delta S_0 [\Psi] }{\delta \Psi(\bphi)} \Bigg{|}_{\Psi_s} = 0\,,
\end{equation}
from which follows the saddle-point equation
\begin{eqnarray}
\Psi_s(\bphi) &=& \frac{ H_z( \bphi)    
\Big(  \int d \bpsi \, U(\bphi,\bpsi )  \Psi_s(\bpsi)    \Big)^{c-1}}
{ \int d \bphi  H_z( \bphi)    
\Big(  \int d \bpsi \, U(\bphi,\bpsi )  \Psi_s(\bpsi)    \Big)^c}\,, \nonumber \\
&=& r_s(\bphi)  \int d \bpsi \, U(\bphi,\bpsi )  \Psi_s(\bpsi) 
\,.
\label{saddle}
\end{eqnarray}
In order to extract the $O(1/N)$ correction to the distribution
of eigenvalues we need to consider the effect of finite size fluctuations in $\Psi_s(\bphi)$.
The full action $S^{(N)} [\Psi]$ can be formally expanded around $\Psi_s(\bphi)$ as follows
\begin{align}
S^{(N)} [\Psi] &= S^{(N)} [\Psi_s] + \int d \bphi \, \frac{\delta S^{(N)} [\Psi] }{\delta \Psi(\bphi)} \Bigg{|}_{\Psi_s}
\left[ \Psi(\bphi) - \Psi_s(\bphi) \right]  \nonumber \\
&+ \frac{1}{2} \int d \bphi \, d \bpsi \,  \frac{\delta^2 S^{(N)} [\Psi] }{\delta \Psi(\bphi) \delta \Psi(\bpsi)} \Bigg{|}_{\Psi_s} \nonumber \\
&\times
\left[ \Psi(\bphi) - \Psi_s(\bphi) \right]  \left[ \Psi(\bpsi) - \Psi_s(\bpsi) \right]  \,.
\label{act}
\end{align}
Assuming that the deviations 
from $\Psi_s(\bphi)$ are of $O(1/\sqrt{N})$ and retaining terms up
to order $O(1/N)$ in the above expansion, we substitute eq. (\ref{act}) in 
eq. (\ref{saddint}) and integrate over
the Gaussian fluctuations to obtain
\begin{equation}
\left\langle \left[ Z_N(z) \right]^n  \right\rangle = \frac{\sqrt{\det{\left( c \, \bU \right)} }}{\sqrt{\det \bJ_0} }
\exp{\left( - N S_0 [\Psi_s] - S_1 [\Psi_s]  \right)} \,,
\label{lpcs}
\end{equation}
where eq. (\ref{action1}) has been used. The elements of $\bJ_0$ read
\begin{equation}
J_0(\bphi,\bpsi) = \frac{\delta^2 S_0 [\Psi] }{\delta \Psi(\bphi) \delta \Psi(\bpsi)} \Bigg{|}_{\Psi_s} \,.
\label{lpq}
\end{equation}
The explicit computation of the derivatives in eq. (\ref{lpq}) and the subsequent
use of eq. (\ref{saddle}) leads to the following expression for $\bJ_0$
\begin{equation}
\bJ_0 = c \, \bU - c \, \bU \bT \,,
\label{dew}
\end{equation}
where we have introduced the matrices
\begin{equation}
T(\bphi,\bpsi) = (c-1) M(\bphi,\bpsi) - c  \int d \bpsi^{\prime} U(\bpsi,\bpsi^{\prime}) \Psi_s (\bphi) \Psi_s (\bpsi^{\prime}) 
\end{equation}
and
\begin{equation}
M(\bphi,\bpsi)= U(\bphi,\bpsi) r_s(\bphi) \,.
\end{equation}
By inserting eq. (\ref{dew}) in eq. (\ref{lpcs}) and employing the
identity $\ln \det{\bX} = {\rm Tr} \ln \bX$ (here $\bX$ denotes a generic
matrix), we obtain the expression 
\begin{equation}
\frac{1}{N} \ln \left\langle \left[ Z_N(z) \right]^n  \right\rangle = 
- S_0 [\Psi_s] - \frac{1}{N} S_1 [\Psi_s] + \frac{1}{N} \sum_{L=1}^{\infty} \frac{{\rm Tr} \bT^L }{2L} \,.
\label{eqZ}
\end{equation}
By substituting eq. (\ref{eqZ}) in eq. (\ref{rhorepl}) and noting 
that the following identity holds
\begin{equation}
{\rm Tr} \bT^L =  (-1)^L + (c-1)^L \left( {\rm Tr} \bM^L -1   \right)\,,
\end{equation}
the first two terms of the
series in eq. (\ref{eqZ}) cancel exactly 
with $S_1 [\Psi_s]$ and we arrive at the following 
expression for $R^{(N)} (z)$
\begin{equation}
R^{(N)} (z) = R_0 (z) + \frac{1}{N} R_1 (z) \,,
\label{auxz}
\end{equation}
where 
\begin{align}
R_0 (z) &= 2 
 \frac{\partial}{\partial z} \lim_{n \rightarrow 0} \frac{\partial}{\partial n}  
 S_0 [\Psi_s]  \,, \label{rho0}  \\ 
R_1 (z) &= 2 
 \frac{\partial}{\partial z} \lim_{n \rightarrow 0} \frac{\partial}{\partial n}  
  \sum_{L=3}^{\infty} \frac{(c-1)^L}{2L} \left( 1 - {\rm Tr} \bM^L   \right) . 
\label{rho1}
\end{align}
This formula should be compared to similar formulae in \cite{Carlo2013,Carlo2014}. Substituting eq. (\ref{auxz}) 
in eq. (\ref{Dosfull}), we
obtain the leading term $\rho_0(\lambda) $ and the $O(1/N)$ 
correction $\rho_1(\lambda) $
to the eigenvalue distribution:
\begin{equation}
\rho_0(\lambda) = \frac{1}{\pi} \lim_{\eta \rightarrow 0^{+}} {\rm Im} \left[ R_0(z) \right] \,, \ \ 
\rho_1(\lambda)  = \frac{1}{\pi} \lim_{\eta \rightarrow 0^{+}} {\rm Im} \left[ R_1(z) \right].
\label{rhos}
\end{equation}
In the next section we show how the limit $n \rightarrow 0$ is taken 
by assuming a particular form for the saddle-point solution $\Psi_s (\bphi)$.


\section{The distribution of eigenvalues in the replica symmetric theory}

The structure of eq. (\ref{saddle}) suggests that
we seek for a saddle-point solution $\Psi_s (\bphi)$ invariant
under orthogonal transformations.
Indeed, it has been established that 
the replica symmetric (RS) saddle-point, which preserves
both rotational and permutation symmetry in the
replica space, yields exact results
for the eigenvalue distribution of several sparse random graph models
\cite{RB88,Dean,K08,Kuhn2009,Tim2010,Kuhn2011,Nagao2013}. In 
particular, the correct analytical expression for $\lim_{N \rightarrow \infty} \rho^{(N)} (\lambda)$ 
in the case of regular random graphs is recovered by the
RS solution. These results are also confirmed by 
reference \cite{bordenave}, where the exactness of the RS assumption is 
proved rigorously for a large class of sparse random graphs with arbitrary
degree distributions.

We thus assume that $\Psi_s (\bphi)$ is an uncountable superposition of
Gaussians \cite{Dean,K08} 
\begin{equation}
\Psi_s (\bphi) = \frac{1}{\mathcal{F}(n)} \int d g Q(g) \prod_{\alpha=1}^{n} \left(\frac{i}{2 \pi g} \right)^{\frac{1}{2}} 
\exp{\left(\frac{- i \phi_{\alpha}^2}{2 g}\right)} \,,
\label{RS}
\end{equation}
where $Q(g)$ is the normalized distribution of the
complex variance $g$ with ${\rm Im} \, g > 0$, such that
the above integral is convergent. The factor $\mathcal{F}(n)$ accounts for
the fact that $\Psi_s (\bphi)$ is not normalized for arbitrary $n$, as can
be noted from eq. (\ref{saddle}). Plugging eq. (\ref{RS})
into eq. (\ref{saddle}) and integrating over $\bphi$, one can determine $\mathcal{F}(n)$ up
to order $O(n)$
\begin{eqnarray}
\left[ \mathcal{F}(n) \right]^2 &=& 1 + \frac{n}{2} \int dg W(g) \ln{\left( \frac{2 \pi g}{i}  \right)}  \nonumber \\
&-&  \frac{n}{2} \int dg Q(g) \ln{\left( \frac{2 \pi g}{i}  \right)}  \,,
\end{eqnarray}
and, in addition, the self-consistent 
equations for
the distributions $Q(g)$ and $W(g)$ 
\begin{align}
Q(g) = \int \left( \prod_{k=1}^{c-1} d g_k Q(g_k) \right) \delta\left( g - \frac{1}{z - \sum_{k=1}^{c-1} g_k }   \right), \label{eqQ} \\
W(g) = \int \left( \prod_{k=1}^{c} d g_k Q(g_k) \right) \delta\left( g - \frac{1}{z - \sum_{k=1}^c g_k }   \right). \label{eqW}  
\end{align}
Equations (\ref{eqQ}) and (\ref{eqW}) can be also derived
trough the more intuitive cavity method, where a clear physical interpretation emerges \cite{Metz2010}. 
The function $W(g)$ is the distribution  
of $\{ \bG_{ii}(z) \}_{i=1,\dots,N}$, while $Q(g)$ is the distribution
of the diagonal elements of $\bG(z)$ on the cavity graph, namely, a graph
where a randomly chosen vertex and all its edges are removed. 
It is straightforward to check that 
$Q(g) = \delta\left( g - g_c  \right)$ and $W(g) = \delta\left[ g - \left( z - c \, g_c \right)^{-1}  \right]\,$
solve, respectively, eqs. (\ref{eqQ}) and (\ref{eqW}), with $g_c$
denoting one of the roots of the quadratic equation
\begin{equation}
(c-1) g_c^2 - z g_c + 1 = 0 \,.
\label{quadr}
\end{equation}
The fact that $Q(g)$ and $W(g)$ are delta peak distributions simply
reflects the absence of fluctuations on the degrees and on the edges
of the graph.

One needs to be careful in choosing the root of 
eq. (\ref{quadr}) depending on the value of $z$.
The natural choice for $g_c$ is the following
\begin{align}
g_c &= \left\{ \begin{array}{ccc}
\frac{1}{2(c-1)} \left(z + \sqrt{z^2 - \lambda_b^2}     \right)  &  & {\rm if} \,\, |z| < |\lambda_b| \\ 
\frac{1}{2(c-1)} \left(z - \sqrt{z^2 - \lambda_b^2}     \right) &  & {\rm if} \,\, |z| \geq |\lambda_b|   \end{array}
\right. ,
\label{casesgc}
\end{align}
where $|\lambda_b| = 2 \sqrt{c-1}$. 
Equation (\ref{casesgc}) ensures that the leading term of the resolvent $R_0(z)$ is an analytic
function of $z= \lambda - i \eta$.
Besides that, this choice for $g_c$ reproduces the correct physical behavior
$R_0(z) = 1/z$ for
$|z| \rightarrow \infty$, since $g_c \rightarrow 0$ in this case \footnote{
For $z = \lambda < 0$ and $|\lambda| \geq |\lambda_b|$, we need to make
the replacement $\sqrt{\lambda^2 - \lambda_b^2} \rightarrow  -\sqrt{\lambda^2 - \lambda_b^2}$
in order to obtain that $g_c \rightarrow 0$ for $\lambda \rightarrow -\infty$ 
and, consequently, derive the correct behavior of $R_0(\lambda)$ and $R_1(\lambda)$ 
in this regime.}.
This decay of $R_0(z)$ implies in the normalization $\int d \lambda \, \rho_0(\lambda) = 1$, as
can be noted from the Stieltjes transform of $\rho^{(N)} (\lambda)$.

Inserting the RS {\it ansatz} for $\Psi_s (\bphi)$ in
eq. (\ref{S0}) and taking the limit $n \rightarrow 0$, an
analytical expression for $R_0(z)$ is derived through eq. (\ref{rho0}). 
For $\eta \rightarrow 0^+$, $R_0(z)$ has a nonzero imaginary
part only if $|\lambda| < |\lambda_b|$, from which the KM law follows 
using eq. (\ref{rhos})
\begin{align}
\rho_0(\lambda) &= \left\{ \begin{array}{ccc}
 \frac{c}{2 \pi} \frac{\sqrt{\lambda_b^2 - \lambda^2} }{\left( c^2 - \lambda^2 \right)}    &  & {\rm for} \,\, |\lambda| < |\lambda_b|  \\ 
  0  &  & {\rm for} \,\, |\lambda| \geq |\lambda_b|  \end{array}
\right. . \label{KMfg}
\end{align}

For the calculation of $R_1(z)$ one needs
to obtain the RS form of $r_s(\bphi)$. This is
achieved by substituting
eq. (\ref{RS}) in eq. (\ref{eqr}) and expanding the result
up to order $O(n)$
\begin{align} 
r_s (\bphi) = \left[1 - \frac{n}{2} \ln{\left( \frac{2 \pi g_c}{i} \right) }   \right]
\exp{\left[ \frac{i \bphi^2}{2} \Big( (c-2) g_c - z  \Big)   \right]}\,,
\end{align}
which allows us to perform the limit $n \rightarrow 0$ 
in eq. (\ref{rho1}) and derive the expression:
\begin{equation}
 R_1 (z) =  
  \sum_{L=3}^{\infty} \frac{(c-1)^L}{2L} \frac{\partial}{\partial z}   \left( L \ln{g_c}    - 2 \ln{Z_L^{(c)}(g_c) }   \right)  \,. 
\label{rho1A}
\end{equation}
The object $Z_L^{(c)}(g_c)$ , defined analogously to eq. (\ref{generat}), is the generating function 
associated to the $L \times L$ tridiagonal matrix $\mathcal{H}$, whose elements are given by
\begin{equation}
\mathcal{H}_{ij} = (c-2) \, g_c \, \delta_{ij} + \delta_{i,j-1} + \delta_{i,j+1} \,, \quad i + N \equiv i\,.
\label{matrH}
\end{equation}
The physical meaning of eq. (\ref{rho1A})
is quite transparent. The object $\frac{\partial}{\partial z}  \ln{Z_L^{(c)}(g_c) }$
can be seen as the resolvent of a 1D closed chain or 
loop of length $L$, where
each node receives an effective field $g_c$ from each one
of its $(c-2)$ neighbors living outside the loop. 
We point out that, at the level of the $O(1/N)$
correction, each node belongs only to a single
loop, i.e., there are no intersecting loops, since these
objects arise on average in a fraction $O(1/N^2)$ of nodes.
The quantity $\frac{\partial}{\partial z} \ln g_c$ is the
resolvent of a 1D closed chain of infinite length \cite{Economoubook1}.
As a consequence, the $O(1/N)$ fluctuations due to all loops 
of a certain length $L$ modify $R^{(N)} (z)$ by a term proportional on average to the difference
between the resolvent of an infinite loop and the resolvent of a finite
loop of length $L$. The weight $\frac{(c-1)^L}{2L}$ is the
average number of loops of length $L$ in a regular random
graph of degree $c$ \cite{Bollobas80,Wormald81}. 
A result analogous to eq. (\ref{rho1A}) has been derived 
in the study of the
$O(1/N)$ corrections to the free-energy of disordered spin
systems defined  on sparse random graphs \cite{Carlo2013}. 

The Gaussian integral in $Z_L^{(c)}(g_c)$ is evaluated using the
eigenvalues of the matrix $\mathcal{H}$, given by $a_n=g_c(c-2)+2 \cos{\left( 2\pi n/L \right)}$, $n=0,\ldots,L-1$, which allows us to
compute in eq. (\ref{rho1A}) the derivative 
with respect to $z$ 
\begin{align}
 R_1 (z) &= \sum_{L=3}^{\infty} \frac{(c-1)^L}{2L} \nonumber \\
&\times \left[
\sum_{n=0}^{L-1} \frac{\left( 1 - (c-2) \frac{\partial g_c}{\partial z}  \right)  }
{\left(z - (c-2) g_c - 2 \cos{\left(\frac{2 \pi n}{L} \right)}     \right)}
+ \frac{L}{g_c} \frac{\partial g_c}{\partial z}  \right] .
\end{align}
From now on, the calculation depends, according to eq. (\ref{casesgc}), whether $|z| < |\lambda_b|$
or $|z| \geq |\lambda_b|$, from which the following expression
for $\frac{\partial g_c}{\partial z}$ is obtained
\begin{align}
\frac{\partial g_c}{\partial z}   &= \left\{ \begin{array}{ccc}
\frac{g_c}{\sqrt{z^2 - \lambda_b^2}} &  &  {\rm if} \,\, |z| < |\lambda_b|    \\ 
 -\frac{g_c}{\sqrt{z^2 - \lambda_b^2}}   &  &  {\rm if} \,\, |z| \geq |\lambda_b|   \end{array}
\right. .
\end{align}
This leads to the following simplified form of $R_1(z)$ 
\begin{align}
R_1 (z) &=  \frac{{\rm sign}\left(|z| - |\lambda_b| \right)}{2 \sqrt{z^2 - \lambda_b^2 }} \sum_{L=3}^{\infty} (c-1)^L  \nonumber \\
&\times
\Bigg{\{} \frac{\left[ (c-2) z +  c  B(z)    \right]}{2 \pi} \mathcal{G}_L(z) - 1  \Bigg{\}},
\label{rho1Ab}
\end{align}
where we have defined
\begin{align}
 \mathcal{G}_L(z)  &= \frac{2 \pi}{L} \sum_{n=0}^{L-1} F_z(x_n) \,,  \quad x_n = \frac{2 \pi n}{L} \,, \label{GLa} \\
B(z) &= {\rm sign}\left(|z| - |\lambda_b| \right)  \sqrt{z^2 - \lambda_b^2}\,,
\end{align}
with
\begin{equation}
F_z(x)  = \frac{1}{c z +  (c-2) B(z) - 4(c-1) 
\cos{(x)}  }  
\end{equation}
and ${\rm sign}(0) \equiv 1$. In the limit $L \rightarrow \infty$,  $\mathcal{G}_L(z)$ becomes simply an
integral of the periodic function $F_z(x)$, which is solved using standard contour integration methods. The result reads
\begin{equation}
\lim_{L \rightarrow \infty} \mathcal{G}_L(z) =   \int_{0}^{2 \pi} d x F_z(x) =   \frac{2 \pi}{(c-2) z 
+  c B(z)   } \,.
\label{integr}
\end{equation}
It follows that the individual terms of the loop series in eq. (\ref{rho1Ab}) are composed
of the exponential growing factor $(c-1)^L$ multiplied by a
function that is going to zero for $L \rightarrow \infty$. The
key point consists in understanding how fast this function 
vanishes as a function of $L$. We will see in the next section that one
can extract the explicit dependence of the summands with
respect to $L$ by borrowing techniques used to compute
the discretization error in the trapezoidal method of numerical integration.


\section{The loop series and the final expression for $\rho_1(\lambda)$}

The problem of studying how $\mathcal{G}_L(z)$ approaches its
asymptotic form $\lim_{L \rightarrow \infty} \mathcal{G}_L(z)$ is
equivalent to evaluate the error
of replacing the sum in eq. (\ref{GLa}) by the integral of eq. (\ref{integr}).
This is analogous to compute the discretization error
in some numerical integration methods, where several techniques 
are available \cite{Philip}. Here we extract the dependence
of  $\mathcal{G}_L(z)$ with respect to $L$
via a Fourier analysis, following steps typically
employed to compute the discretization error in the trapezoidal
rule of numerical integration \cite{Philip,Waldvogel2011,Loyd}. 

Let us expand $F_z(x)$ in a Fourier series
\begin{align}
F_z(x)  &= \frac{a_0}{2} + \sum_{k=1}^{\infty} a_k \cos{(kx)}\,,  \\
 a_k &= \frac{1}{\pi} \int_{-\pi}^{\pi} d x \cos{(k x)} F_z(x) \,, 
\label{fourier}
\end{align}
and assume that this series converges at 
the points $x_n$ ($n=0,\dots,L-1$) defined
in eq. (\ref{GLa}). Plugging  the above expansion 
into $\mathcal{G}_L(z)$ and noting 
that $\lim_{L \rightarrow \infty} \mathcal{G}_L(z) = \pi a_0$, we obtain
an exact equation for the deviation of $\mathcal{G}_L(z)$ with
respect to its $L \rightarrow \infty$ limit
\begin{equation}
\mathcal{G}_L(z) - \lim_{L \rightarrow \infty} \mathcal{G}_L(z) =
2 \sum_{m=1}^{\infty} \int_{-\pi}^{\pi} d x \cos{(m L x)} F_z(x) \,.
\label{fourier1}
\end{equation}
The asymptotic behavior of $\mathcal{G}_L(z)$  is governed by 
the convergence rate of the Fourier series for $F_z(x)$, in full
analogy with the error formula for the trapezoidal 
quadrature \cite{Waldvogel2011,Loyd}. 
In order to make further progress, eq. (\ref{fourier1}) is
substituted
in eq. (\ref{rho1Ab}) and the above integral over $F_z(x)$ is transformed in 
a contour integral along the unit circle in the complex 
plane, traversed once in the counterclockwise direction
\begin{align}
R_1 (z) &=  \frac{-\left[ {\rm sign}\left( |z| - |\lambda_b| \right) (c-2) z + c \sqrt{z^2 - \lambda_b^2}    \right] }
{4 \pi i (c-1) \sqrt{z^2 - \lambda_b^2 }} \nonumber \\
&\times
\sum_{L=3}^{\infty} (c-1)^L 
\sum_{m=1}^{\infty} 
\oint \frac{d \omega \, \omega^{mL}}{\omega^2 + 2 Z_z \, \omega + 1} ,
\label{rho1Ab1}
\end{align}
where
\begin{equation}
 Z_z = - \frac{1}{4 (c-1)} \left[ c z +  (c-2) B(z)    \right] \,.
\label{a2}
\end{equation}

The rest of the analysis amounts to study, in the integrand of eq. (\ref{rho1Ab1}), the behavior
of the poles, i.e., the roots
of the quadratic equation $\omega^2 + 2 Z_z \, \omega + 1 =0$. In
general, one root  $\omega_d$ lies inside the unit circle in the
complex plane, while
the other root $\omega_f$ lies outside. Using
eq. (\ref{casesgc}) and the quadratic equation $g_c = [ z-(c-1) g_c ]^{-1}$, one can
show that $Z_z = - \frac{1}{2} \left(g_c + g_c^{-1} \right)$, from which the
roots $\omega_d$ and $\omega_f$ are computed explicitly
\footnote{We notice {\it en passant} that the matrix elements of the Green function at 
two points separated by a distance $r>0$ are given by $A \left(g_c \right)^r$, with an appropriate 
value of the constant $A$. Therefore, the quantity $g_c$, namely the diagonal part of the Green function
on the cavity graph, is a key quantity of the model: it controls both the large distance decay
of the Green function in the limit $N \rightarrow \infty$ and 
the $1/N$ corrections to the resolvent.}
\begin{equation}
\omega_d = g_c\,, \quad \omega_f = \frac{1}{g_c} \,.
\end{equation}
This allows us to solve the contour integral in eq. (\ref{rho1Ab1}) through the residue theorem and
derive the following expression
\begin{equation}
R_1 (z) = \mathcal{C}(z)  \sum_{L=3}^{\infty} (c-1)^L 
\frac{ g_c^{L}}{1 - g_c^{L} } \,, 
\label{rho1Ab2}
\end{equation}
where the prefactor $\mathcal{C}(z)$ is given by
\begin{align}
\mathcal{C}(z)  &= \left\{ \begin{array}{ccc}
\frac{ \left( z - c \, g_c \right) g_c  }
{ \sqrt{z^2 - \lambda_b^2 } \left( g_c^2 - 1 \right) }  &  &  {\rm if} \,\, |z| < |\lambda_b|    \\ 
  -\frac{ \left( z - c \, g_c \right) g_c  }
{ \sqrt{z^2 - \lambda_b^2 } \left( g_c^2 - 1 \right) }  &  &  {\rm if} \,\, |z| \geq |\lambda_b|   \end{array}
\right. .
\label{casesA}
\end{align}

The $O(1/N)$ correction $R_1 (z)$ to the resolvent  is an analytic function of $z$ with
singularities located possibly only on the real axis. 
In the regime $|z| \rightarrow \infty$, we have that
$g_c = O(1/z)$ and $\mathcal{C}(z)= O(1/z)$, such that
the loop series in eq. (\ref{rho1Ab2}) also converges to zero for large $z$.
It follows that
$R_1 (z)$ vanishes faster than $1/z$, which implies 
that $\int d \lambda \, \rho_1(\lambda) =0$, as can be checked
using the Stieltjes transform of $\rho^{(N)}(\lambda)$.
This is consistent with the normalization of both the full eigenvalue
distribution $\rho^{(N)}(\lambda)$ and its leading term 
$\rho_0(\lambda)$. In the sequel 
we study, separately in the sectors $|z| \geq |\lambda_b|$ and  $|z| < |\lambda_b|$, the behavior of $R_1 (z)$ 
as $\eta \rightarrow 0^{+}$.


\subsection{$|\lambda| \geq |\lambda_b|$: the isolated eigenvalue}

The idea now consists in setting $z = \lambda$ and then
making an analytical continuation from $\lambda \rightarrow \infty$,
where $R_1 (\lambda)$ is convergent, to smaller values of $\lambda$.
In the regime $|\lambda| \geq |\lambda_b|$, the quantity $g_c$ reads
\begin{equation}
g_c = \frac{1}{2(c-1)} \left( \lambda - {\rm sign}(\lambda) \sqrt{\lambda^2 - \lambda_b^2}  \right)\,.
\label{eor}
\end{equation}
One can check that, for $\lambda > c$, $g_c$
fulfills $0 < g_c < 1/(c-1)$ and the loop series in eq. (\ref{rho1Ab2})
is convergent.
For $\lambda =c$, we have that $g_c= 1/(c-1)$, and
the loop series in eq. (\ref{rho1Ab2}) becomes divergent.
This singular behavior is consistent with the existence
of an isolated eigenvalue, located at $\lambda = c$, outside of the 
support $(-|\lambda_b|,|\lambda_b|)$ of the continuous part
of the spectrum. Indeed, for this simple model of RRGs, this isolated
eigenvalue can be computed directly from the eigenvalue
equation and it corresponds to the uniform eigenvector.

For $|\lambda_b| < \lambda < c$, we have that $1/(c-1) < g_c < 1/\sqrt{c-1}$ and
the loop series of eq. (\ref{rho1Ab2}) is divergent. However, we can rewrite this series
as follows
\begin{equation}
\sum_{L=3}^{\infty} (c-1)^L 
\frac{ g_c^{L}}{1 - g_c^{L} } = \sum_{L=3}^{\infty} (c-1)^L g_c^{L} + \sum_{L=3}^{\infty} (c-1)^L 
\frac{ g_c^{2 L}}{1 - g_c^{L} } \,.
\label{eqaux35}
\end{equation} 
The second term on the right hand side is a convergent
series, while we can 
assign a finite
value for the summation of the first term using the standard
expression for the geometric series \cite{hardy1991}, leading
to a finite result for $R_1 (\lambda)$ in 
the range $|\lambda_b| < \lambda < c$.
Finally, we have that $g_c = 1/\sqrt{c-1}$ for $\lambda=|\lambda_b|$
and a second singularity arises, which
corresponds to the edge of the continuous band. 

For a given point $\lambda$ in the regime $\lambda < -|\lambda_b|$, $g_c$ is given by minus 
its value at $|\lambda|$.
Thus the qualitative behavior of  $R_1 (\lambda)$
for $\lambda < -|\lambda_b|$
is completely analogous to the case $\lambda > |\lambda_b|$, with
the exception that $R_1 (\lambda)$ is finite for $\lambda= -c$, since
the first term on the right hand side of eq. (\ref{eqaux35})
is an alternating divergent series that can be summed 
using the summation formula for the geometric series \cite{hardy1991}.
Consequently, $R_1 (\lambda)$
remains finite in the whole sector $\lambda < -|\lambda_b|$, exhibiting a
singularity only at $\lambda = -|\lambda_b|$.
We point out that, according to eq. (\ref{eor}), $g_c \in \mathbb{R}$  
for $|\lambda| \geq |\lambda_b|$. This implies that, for the different sectors 
of $\lambda$ where $R_1 (\lambda)$ attains a finite value, we have that 
$\rho_1 (\lambda) = 0$, since $R_1 (\lambda)$ is also a real-valued function (see eqs. (\ref{rho1Ab2})
and (\ref{casesA})).


\subsection{$|\lambda| < |\lambda_b|$: the continuous band of eigenvalues}

For $\eta \rightarrow 0^+$ and $|\lambda| < |\lambda_b|$, $g_c$ is
obtained from eq. (\ref{casesgc})
\begin{equation}
g_c= \frac{1}{2(c-1)} \left( \lambda + i \sqrt{\lambda_b^2 - \lambda^2}  \right) \,. 
\end{equation}
Inserting the above form of $g_c$ in eq. (\ref{casesA}), one can show
that ${\rm Re} \, \mathcal{C}(\lambda) = 0$. Thus, by taking the imaginary
part of eq. (\ref{rho1Ab2}), the following
expression is derived for the $O(1/N)$ correction $\rho_1 (\lambda)$ to the continuous
part of the eigenvalue distribution
\begin{equation}
\rho_1 (\lambda) = C(\lambda) \, {\rm Re} \left[ \sum_{L=3}^{\infty} (c-1)^L  \frac{g_c^{L}}{1 - g_c^{L}}        \right] \,,
\label{rhofim}
\end{equation}
with 
\begin{equation}
C(\lambda) = \frac{1}{\pi \sqrt{\lambda_b^2 - \lambda^2}} \,.
\label{Cdef1}
\end{equation}
Equation (\ref{rhofim}) can be derived from 
the average of the finite size fluctuations of the
linear eigenvalue functional defined in reference 
\cite{Johnson2012}. The correspondence between $\rho_1(\lambda)$
and the rigorous results of \cite{Johnson2012} is discussed
in appendix \ref{app2}.

The last step consists in handling the loop 
series in eq. (\ref{rhofim}), which is irremediably divergent
since $|g_c| = 1/\sqrt{c-1}$. However, we
can rewrite this series according to
\begin{align}
 \sum_{L=3}^{\infty} (c-1)^L  \frac{g_c^{L}}{1 - g_c^{L}} &= \sum_{L=3}^{\infty} (c-1)^L g_c^{L} + \sum_{L=3}^{\infty} (c-1)^L g_c^{2L} \nonumber \\
&+   \sum_{L=3}^{\infty} (c-1)^L \frac{g_c^{3L}}{\left( 1 - g_c^{L} \right) } \,,
\label{quasifinal}
\end{align}
and, despite the fact that the first two terms on the right hand side are divergent, they
can be summed using the summation formula for the geometric series \cite{hardy1991}.
The series containing $g_c^{3L}$ is
clearly convergent and, in this way, we arrive at the final
expression for $\rho_1 (\lambda)$
\begin{align}
\rho_1 (\lambda) &= C(\lambda) \, {\rm Re} \left[ \frac{(c-1) g_c}{1 - (c-1) g_c } +  \frac{(c-1) g_c^2}{1 - (c-1) g_c^2 } \right] \nonumber \\
&+ C(\lambda) \, {\rm Re} \left[
 \sum_{L=3}^{\infty} (c-1)^L \frac{g_c^{3L}}{\left( 1 - g_c^{L} \right) } - K(g_c)   \right] \,,
\label{final}
\end{align}
where the factor $K(g_c)$ accounts for the absence 
of the terms with $L=1$ and $L=2$ in eq (\ref{quasifinal}):
\begin{equation}
K(g_c) = (c-1) g_c + c(c-1) g_c^2 + (c-1)^2 g_c^4 \,.
\end{equation}
Equation (\ref{final}) constitutes the central result
of this work: it provides the analytical expression
for the $O(1/N)$ correction to the KM distribution for $|\lambda| < |\lambda_b|$.

There is one important point as far as the behavior near $\pm |\lambda_b|$ is concerned.  
In the limit $\lambda \rightarrow \pm |\lambda_b|$, we have  that $C(\lambda)$ 
diverges as $O\left((|\lambda_b| \mp \lambda)^{-1/2}\right)$, while the real part
of the loop series in eq. (\ref{final}) is numerically shown to converge to a negative 
finite value. Thus $ \rho_1(\lambda)$ is a distribution with integrable singularities 
at $\lambda= \pm |\lambda_b|$. There is also a contribution proportional to $\delta(\lambda\pm |\lambda_b|)$ 
because the resolvent has poles at  these points. The details of the 
behavior at the band edges will not be investigated here.

\begin{figure}[h!]
\includegraphics[scale=0.92]{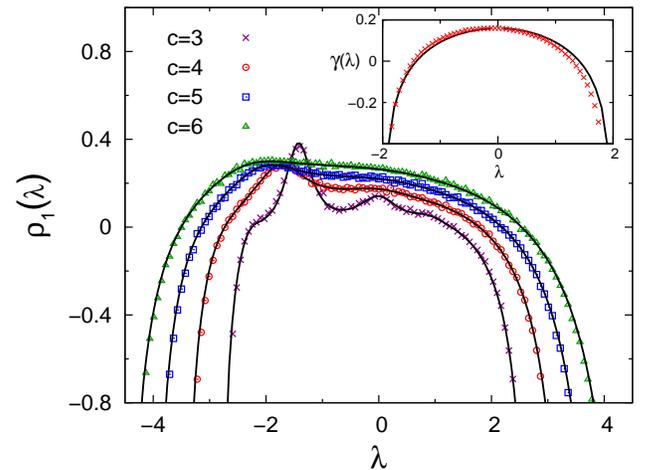}
\caption{The $O(1/N)$ correction to the average eigenvalue distribution 
of the adjacency matrix of an ensemble of regular
random graphs with degree $c$, where the isolated eigenvalue $\lambda=c$
has been omitted. In the main graph, the solid black curves depict
the analytical result of eq. (\ref{final}) for different $c$, while the symbols
represent numerical diagonalization results obtained from matrices of size $N=500$. In 
the inset, the solid black line shows 
the analytical expression for $c \gg 1$, given by eq. (\ref{cases1}), and
the red symbols are direct diagonalization results for $c=40$ and $N=500$. 
The histograms from numerical diagonalizations are
obtained by averaging the results over $5 \times 10^6$
samples. 
}
\label{fig5}
\end{figure}

In figure \ref{fig5} we compare eq. (\ref{final}) with direct diagonalization results of the adjacency matrix 
of regular random graphs with $N=500$, generated according to the algorithm 
presented in reference \cite{WormaldA}. The agreement between theoretical and numerical
results is excellent. For finite $N$, the regular graph becomes sensibly non-bipartite
due to the presence of loops, which is reflected in the
breaking of the symmetry $\lambda \rightarrow -\lambda$ in $\rho_1 (\lambda)$.

After rescaling the adjacency matrix elements as $A_{ij} \rightarrow \frac{A_{ij}}{\sqrt{c-1}}$, one
can show that, in the regime $1 \ll c \ll N$, the dominant contribution
to $\rho_1 (\lambda)$ is given by $\rho_1 (\lambda) = c \, \gamma(\lambda)$, where
the coefficient $\gamma(\lambda)$ reads
\begin{equation}
\gamma(\lambda) = \frac{2 - \lambda^2}{2 \pi \sqrt{4 - \lambda^2}} \,.
\label{cases1}
\end{equation}
The numerical diagonalization results converge for large
$c$ to eq. (\ref{cases1}), as illustrated in the inset
of figure \ref{fig5}.
Although the leading term $\rho_0(\lambda)$ converges to the 
Wigner semicircle law for $c \gg 1$, this is not the
case for the $O(1/N)$ fluctuations, as can be seen by comparing eq. (\ref{cases1})
with the corresponding results in references \cite{Verb84,Dhesi90}.


\section{Final remarks}

The average eigenvalue distribution of a regular
random graph with $N$ vertices converges, in the
limit $N \rightarrow \infty$, to the well-known 
Kesten-McKay (KM) law. 
In this work we have derived an
exact analytical expression for the $O(1/N)$ correction
to the KM law using the replica approach for random matrices. 
The $O(1/N)$ correction is incorporated
in the replica scheme by taking into account the $O(1/\sqrt{N})$ fluctuations
around the mean-field saddle-point solution. Although the intermediate
steps in the replica method are not very
intuitive, the interpretation of the final expression for the $O(1/N)$
fluctuations of the resolvent, cf. eq. (\ref{rho1A}), from
which follows our analytical result, given by eq. (\ref{final}), is
rather clear: it consists of a sum over the average
contributions coming from loops of all possible lengths, each
loop of finite length contributing with a term proportional to the deviation  
of its effective resolvent with respect to the resolvent of an infinite loop.
The approach discussed in this work is also capable to determine
the isolated eigenvalue, since the latter has a weight of $O(1/N)$
in the average eigenvalue distribution.

The ideas presented here can be possibly extended to more
general random graph
models including disordered edges and fluctuating 
connectivities, which opens the possibility to analyze, for
instance, finite size fluctuations in the Anderson model
on the Bethe lattice \cite{Abouchacra73}. Despite the
non-critical behavior of the average density of states along the localization
transition, the study of finite size corrections in such
mean-field models may provide some valuable insights on
the influence of loops in the electronic properties of
finite dimensional models. Besides that, the study
of finite size corrections to the density of states
can be considered as a warm up to the more complicated
task of considering relevant quantities to the localization
transition, such as the inverse participation ratio.

On the methodological side, a derivation of eq. (\ref{rho1A})  
through the cavity method would be a meaningful exercise, since the
latter approach, being conceptually simpler, usually provides additional 
physical insights, which are obscured by the replica calculation. Work along some 
of these lines is underway, following the lines of \cite{Carlo2013,Carlo2014}.

Finally, it would be also interesting to examine
the universality status of the level correlation function
in the case of sparse random graph models \cite{fyodorov91}, using the 
ideas presented in this paper. 
\\

\acknowledgements

GP thanks G\'erard Ben Arous for fruitful discussions.
FLM thanks Carlo Lucibello for interesting comments.
The research leading to these results has received funding from the European Research Council
(ERC) grant agreement No. 247328 (CriPheRaSy project),
from  the People Programme (Marie Curie Actions) of the European Union's Seventh Framework 
Programme FP7/2007-2013/ under REA grant agreement No. 290038  (NETADIS project) and 
from the Italian MIUR under the Basic
Research Investigation Fund FIRB2008 program, grant
No. RBFR08M3P4, and under the PRIN2010 program, grant code 2010HXAW77-008. \\


\appendix

\section{Derivation of the saddle-point integral} \label{app1}

The purpose of this appendix is to discuss the main steps 
involved in the derivation of eq. (\ref{saddint}). 
The average over the topological disorder in eq. (\ref{Zn1})
is calculated using integral representations
for the Kronecker deltas in the distribution $p(\{ A_{i < j} \})$, leading
to 
\begin{widetext}
\begin{align}
\left\langle \left[ Z_N(z) \right]^n  \right\rangle &= \frac{1}{\mathcal{A}_N}
\int \left(  \prod_{i=1}^{N} d \bphi_i H_z(\bphi_i) \right) 
 \int_{0}^{2 \pi} \left(  \prod_{i=1}^{N} \frac{d x_i}{2 \pi} e^{i \, c \, x_i}  \right) 
\exp{\left[   \frac{1}{2} \sum_{ij=1}^{N} \ln{\left[1 + \frac{c}{N} \left(  e^{-i \, ( x_i + x_j)  } U (\bphi_i,\bphi_j) -1     \right) \right]}
\right]} \nonumber \\
&\times
\exp{\left[  - \frac{1}{2} \sum_{i=1}^{N} \ln{\left[1 + \frac{c}{N} \left(  e^{-2i x_i  } U (\bphi_i,\bphi_i) -1     \right) \right]}
\right]}\,, \nonumber
\end{align}
\end{widetext}
where $U (\bphi,\bpsi) = \exp{\left( i \bphi.\bpsi \right)}$ and $H_z(\bphi)$ is defined
by eq. (\ref{Hdef}). Since we are interested in the
$O(1/N)$ correction to the average spectrum, we need to determine the exponent of the 
above integrand up to $O(N^0)$. After performing an expansion 
in powers of
$1/N$, the sites are decoupled via the introduction, by means of the
Fourier integral representation of the Dirac delta, of the functional order-parameters
\begin{align}
\mu_1 (\bphi) &= \frac{1}{N} \sum_{i=1}^{N} \delta(\bphi - \bphi_i) e^{- i x_i} \,, \nonumber \\
\qquad \mu_2 (\bphi) &= \frac{1}{N} \sum_{i=1}^{N} \delta(\bphi - \bphi_i) e^{- 2 i x_i}\,, \nonumber
\end{align}
which allows us to recast $\left\langle \left[ Z_N(z) \right]^n  \right\rangle$ in the
form
\begin{align}
&\left\langle \left[ Z_N(z)  \right]^n   \right\rangle = \frac{\exp{\left(-\frac{Nc}{2} - \frac{c^2}{4} + \frac{c}{2}  \right)}}{\mathcal{A}_N} 
\int \mathcal{D} \mu_1 \mathcal{D} \hat{\mu}_1 \mathcal{D} \mu_2 \mathcal{D} \hat{\mu}_2 \nonumber \\
&\times
\exp{\left[ i \int d \bphi  \left[ \mu_1(\bphi)  \hat{\mu}_1(\bphi)  +  \mu_2(\bphi)  \hat{\mu}_2(\bphi)  \right] \right]} \nonumber \\
&\times
\exp{\left[ N \ln{\mathcal{I}[\hat{\mu_1},\hat{\mu_2}]}  - \frac{c}{2} \int  d \bphi  \mu_2(\bphi) U(\bphi,\bphi)         \right]    }   \nonumber \\
&\times
\exp{\left[  
\frac{c}{2}(N+c) \int  d \bphi    d \bpsi \mu_1(\bphi) U(\bphi,\bpsi) \mu_1(\bpsi)   \right] } \nonumber \\
&\times
\exp{\left[  
- \frac{c^2}{4}  \int  d \bphi d \bpsi   \mu_2(\bphi)   \left( U(\bphi,\bpsi)  \right)^2    \mu_2(\bpsi)       
 \right] } \,,
\label{eqaux}
\end{align}
where
\begin{align}
\mathcal{I}[\hat{\mu_1},\hat{\mu_2}] &=  \int  d \bphi \, H_z(\bphi)  \int_{0}^{2 \pi} \frac{d x}{2 \pi} \nonumber \\
&\times
\exp{\left[i c x - \frac{i}{N} \hat{\mu_1}(\bphi) e^{- i x}  - \frac{i}{N} \hat{\mu_2}(\bphi)   e^{- 2 i x}  \right] }\,.
\label{eqI}
\end{align}
Each integration measure  $\mathcal{D} \mu_1,\dots, \mathcal{D} \hat{\mu}_2$  in eq. (\ref{eqaux}) includes an unimportant factor $1/\sqrt{2 \pi}$
coming from the Fourier representation of the Dirac delta function. 
The integral
over $x$ in eq. (\ref{eqI}) is calculated using the power-series representation
\begin{equation}
\exp{\left[ - \frac{i}{N} \hat{\mu_I}(\bphi) e^{- i I x}   \right]} 
= \sum_{k=0}^{\infty} \left(  - \frac{i}{N} \hat{\mu_{I}}(\bphi)   \right)^{k} \frac{e^{- i I k x}}{k!} \,,
\label{hqpa}
\end{equation}
with $I=1,2$. By substituting eq. (\ref{hqpa}) in eq. (\ref{eqI}) and 
integrating over $x$, we obtain
\begin{align}
\mathcal{I}[\hat{\mu_1},\hat{\mu_2}] &=  \int  d \bphi \, H_z(\bphi) 
\sum_{k=0}^{\infty} \frac{\left[ - i  \hat{\mu_{2}}(\bphi)  \right]^{k}}{ k!}
\frac{\left[ - i  \hat{\mu_{1}}(\bphi)  \right]^{c- 2 k}}{N^{c- k} (c- 2 k)!} .
\label{eqksl}
\end{align}
After performing the rescaling $\hat{\mu}_1(\bphi) \rightarrow N  \hat{\mu}_1(\bphi)$, eq. (\ref{eqksl})
can be expanded up to $O(1/N)$, which yields, after the substitution of the result in eq. (\ref{eqaux}), the following expression 
\begin{widetext}
\begin{align}
\left\langle \left[ Z_N(z) \right]^n  \right\rangle &= \frac{\exp{\left(-\frac{Nc}{2} - \frac{c^2}{4} + \frac{c}{2}  \right)}}{\mathcal{A}_N} 
\int \mathcal{D} \mu_1 \mathcal{D} \hat{\mu}_1 \mathcal{D} \mu_2 \mathcal{D} \hat{\mu}_2
\exp{\left[ i \int d \bphi  \left( N \mu_1(\bphi)  \hat{\mu}_1(\bphi)  +  \mu_2(\bphi)  \hat{\mu}_2(\bphi)  \right) 
\right]    } \nonumber \\
&\times \exp{\left[ \frac{c}{2}(N+c) \int  d \bphi    d \bpsi \mu_1(\bphi) U(\bphi,\bpsi) \mu_1(\bpsi) 
- \frac{c}{2} \int  d \bphi  \mu_2(\bphi) U(\bphi,\bphi)      
+ N \ln{\left( \int \frac{d \bphi}{c!} H_z(\bphi) \left[ - i \hat{\mu}_1(\bphi)  \right]^c           \right) }
     \right]  } \nonumber \\
&\times \exp{\left[ - \frac{c^2}{4}  \int  d \bphi d \bpsi   \mu_2(\bphi)   \left( U(\bphi,\bpsi)  \right)^2    \mu_2(\bpsi)  
+ i \int d \bphi \mathcal{R}[\hat{\mu}_1(\bphi)] \hat{\mu}_2(\bphi)     \right]  } \,,
\nonumber
\end{align}
\end{widetext}
where we have defined
\begin{equation}
\mathcal{R}[\hat{\mu}_1(\bphi)] = c (c-1) \frac{H_z(\bphi) \left[ \hat{\mu}_1(\bphi)    
\right]^{c-2}}{\int d \bphi H_z(\bphi) \left[ \hat{\mu}_1(\bphi)    \right]^{c}} \,,
\end{equation}
and $\mathcal{D} \hat{\mu}_1 = \prod_{ \{ \bphi \} } N/\sqrt{2\pi} \, d \hat{\mu}_1(\bphi)$, while
the other integration measures are defined similarly, but without the factor $N$.
Now one can integrate over $\hat{\mu}_2$, $\mu_2$ and $\mu_1$ to obtain
\begin{align}
&\left\langle \left[ Z_N(z) \right]^n  \right\rangle =
\frac{\exp{\left(-\frac{Nc}{2} - \frac{c^2}{4} + \frac{c}{2}  \right)}}
{\mathcal{A}_N \left[\det{\left( \bU \left( -c - \frac{c^2}{N}  \right)  \right)} \right]^{\frac{1}{2}}} \nonumber \\
&\times
\int \mathcal{D} \hat{\mu}_1 \exp{\left[ \frac{c}{2} \int d \bphi \,  \mathcal{R}[\hat{\mu}_1(\bphi)] U(\bphi,\bphi)  \right]} \nonumber \\
&\times
\exp{\left[ \frac{N}{2 c} \left(1 - \frac{c}{N}  \right)  \int d \bphi d \bpsi  \hat{\mu}_1(\bphi) U^{-1}(\bphi,\bpsi)  \hat{\mu}_1(\bpsi)     
\right]}    \nonumber \\
&\times
\exp{\left[   - \frac{c^2}{4} \int d \bphi d \bpsi \mathcal{R}[\hat{\mu}_1(\bphi)]  \left( U(\bphi,\bpsi)  \right)^2  
\mathcal{R}[\hat{\mu}_1(\bpsi)] \right]} \nonumber \\
&\times
\exp{\left[ N \ln{\left( \int \frac{d \bphi}{c!} H_z(\bphi) \left[ - i \hat{\mu}_1(\bphi)  \right]^c  \right)}    \right]} \,,   
\label{Zfim}
\end{align}
where the integration measure becomes $\mathcal{D} \hat{\mu}_1 = \prod_{ \{ \bphi \} } \sqrt{N/2\pi} \, d \hat{\mu}_1(\bphi)$.
The last step consists in calculating the normalization
factor $\mathcal{A}_N$ from eq. (\ref{distrc})
\begin{align}
\mathcal{A}_N &= \exp{\left[ N \left( - c + c \ln c - \ln c! \right) \right]} \nonumber \\
&\times \exp{ \left[ \frac{c}{2} + \frac{1}{4} - \frac{1}{2} \ln 2 + O\left( \frac{1}{N} \right)  \right]}\,.
\label{normA}
\end{align}
Substituting eq. (\ref{normA}) in eq. (\ref{Zfim})
and making the following change
of the integration variable
\begin{equation}
\hat{\mu}_1(\bphi) = i c \int d \bpsi \, U(\bphi,\bpsi)  \Psi(\bpsi) \,,
\end{equation}
one can rewrite $\left\langle \left[ Z_N(z) \right]^n  \right\rangle$ 
as in eq. (\ref{saddint}).


\section{Correspondence with rigorous results} \label{app2}

The main rigorous result of reference \cite{Johnson2012} is the following 
theorem:

\vspace{0.5cm}

{\it Fix $c \geq 3$ and let $G_N$ be a random $c$-regular graph on $N$ vertices
with adjacency matrix $A_N$. Let $\lambda_1 \geq \dots \geq \lambda_N$ be the eigenvalues 
of $(c-1)^{- 1/2} A_N$.

Suppose that $f$ is a function defined on the complex plane, analytic inside
a Bernstein ellipse of radius $2 \rho$, where $\rho=(c-1)^{\alpha}$ for some $\alpha > 3/2$, and such
that $|f(z)|$ is bounded inside the ellipse. Then $f(x)$ can be expanded on $[-2,2]$ as
\begin{equation}
f(x) = \sum_{k=0}^{\infty} a_k \Gamma_k(x) ,
\label{lpq12}
\end{equation}   
and $Y_f^{(N)} = \sum_{i=1}^N f(\lambda_i) - N a_0$ converges in law as $N \rightarrow \infty$ to the
infinitely divisible random variable
\begin{equation}
Y_f = \sum_{k=1}^{\infty} \frac{a_k}{(c-1)^{k/2}} CNBW_k^{(\infty)} \,.
\label{eqpot}
\end{equation}
}

Let us specify the important quantities that appear in this theorem.
The polynomials $\Gamma_k(x)$ are defined according to
\begin{align}
\Gamma_0(x) &= 1 \,, \label{lrd} \\
\Gamma_{2k}(x) &= 2 T_{2k}\left(\frac{x}{2}\right) + \frac{c-2}{(c-1)^k} \quad k \geq 1 \,, \\
\Gamma_{2k+1}(x) &= 2 T_{2k+1}\left(\frac{x}{2}\right) \quad k \geq 0 \,,
\label{kqp}
\end{align}
where $T_k(x)$ are Chebyshev polynomials of the first kind, which fulfill the
orthogonality relations
\begin{align}
\int_{-1}^{1} \frac{d x}{ \sqrt{1-x^2}} T_i(x) T_j(x)  &= \left\{ \begin{array}{ccc}
0 &  &  {\rm if} \,\,  i \neq j   \\ 
\pi  &  &  {\rm if} \,\, i=j=0 \\
\frac{\pi}{2}  &  &  {\rm if}  \,\, i = j \neq 0
  \end{array}
\right. .
\label{casesww}
\end{align}
The random variable $CNBW_k^{(\infty)}$ is the number of cyclically non-backtracking walks of
length $k$ in $G_N$ \cite{Johnson2012}. It has the explicit form 
\begin{equation}
CNBW_k^{(\infty)} = \sum_{j|k} 2 j C_j^{(\infty)} \,,
\end{equation}
where the sum runs over the values $j=3,\dots,\infty$ such that $\frac{k}{j}$ is an 
integer. The variables $C_j^{(\infty)}$ are independent Poisson random numbers
with average $(c-1)^j/2 j$.

The above theorem makes a statement about the deviation of 
a general linear functional of the
eigenvalues, defined by $\sum_{i=1}^N f(\lambda_i)$, with respect
to the quantity $N a_0$, as $N$ grows to infinity. It tells us
that the deviation $\sum_{i=1}^N f(\lambda_i) - N a_0$ converges in distribution
to a non-Gaussian random variable $Y_f = O(1)$, defined in eq. (\ref{eqpot}). 
Hence we can write down the following equation for the ensemble average of the linear
functional
\begin{equation}
\frac{1}{N} \left\langle \sum_{i=1}^N f(\lambda_i) \right\rangle 
= a_0 +  \frac{1}{N} \sum_{k=1}^{\infty} \frac{a_k}{(c-1)^{k/2}} V_k
\label{kl}
\end{equation}
where
\begin{equation}
V_k = \sum_{j|k} (c-1)^j \,.
\label{lpqde}
\end{equation}
The right hand side of eq. (\ref{kl}) has been obtained by taking the average over the Poisson random variables
present in $CNBW_k^{(\infty)}$. Note also that $V_1=V_2=0$, because the sum over $j$ in the definition
of $V_k$ starts at $j=3$.

We have computed the $O(1/N)$ correction to
the averaged resolvent
\begin{equation}
R_N(z) = \frac{1}{N} \langle {\rm Tr} \bG(z) \rangle = \frac{1}{N} \left\langle \sum_{i=1}^N  \frac{1}{z - \lambda_i} \right\rangle \,,
\end{equation}
with $z = \lambda - i \eta$. Thus, $R_N(z)$ is the ensemble average of a linear
functional of the form $N^{-1} \sum_{i=1}^N f(\lambda_i)$, with $f(\lambda_i) = 1/\left(z -\lambda_i\right)$.
From eq. (\ref{kl}), we expect that $a_0$ gives the leading behavior of $R_N(z)$, while
the other coefficients $a_1,\dots,a_\infty$ contain information about the $O(1/N)$ fluctuations.
Thus, the computation boils down to determine $a_0,\dots,a_\infty$.

From eq. (\ref{lpq12}), we can write
\begin{equation}
f(2 x) = a_0 \Gamma_0 (2 x) + \sum_{k=1}^{\infty} a_{2 k} \Gamma_{2 k} (2 x)+ \sum_{k=0}^{\infty} a_{2 k+1} \Gamma_{2 k+1} (2 x)\,.
\end{equation}
By multiplying both sides by $\Gamma_j(x)/\sqrt{1-x^2}$, integrating over $x$ and using eqs. (\ref{lrd}-\ref{kqp}) and (\ref{casesww}), one 
derives the following expressions for the coefficients
\begin{align}
a_0 &= \frac{1}{\pi} \int_{-1}^{1} \frac{d x}{ \sqrt{1-x^2}} f(2 x) \left[1 - (c-2) \sum_{j=1}^{\infty} \frac{T_{2 j}(x)}{(c-1)^j}   \right] \label{ypq} \\
a_{2 k + 1} &= \frac{1}{\pi} \int_{-1}^{1} \frac{d x}{ \sqrt{1-x^2}} T_{2 k +1}(x) f(2 x) \quad k \geq 0 \label{re4} \\
a_{2 k } &= \frac{1}{\pi} \int_{-1}^{1} \frac{d x}{ \sqrt{1-x^2}} T_{2 k}(x) f(2 x) \quad k \geq 1 \label{re5}
\end{align}

There are many ways to write down an explicit form for the Chebyshev 
polynomials. Here we use the following expression
\begin{equation}
T_j (x) = \frac{1}{2} \left[ x + i \sqrt{1 - x^2}   \right]^{j} + \frac{1}{2} \left[ x - i \sqrt{1 - x^2}   \right]^{j} \,,
\label{laqp}
\end{equation}
valid in the domain $x \in [-1,1]$.
By substituting eq. (\ref{laqp}) in eq. (\ref{ypq}), we get
\begin{equation}
a_0 =  \frac{  2 c \, (c-1)  }{\pi} \int_{-1}^{1} d x \, f(2 x) \frac{ \sqrt{1-x^2}}{ \left[ c^2 - 4 x^2 (c-1) \right]} \,.
\end{equation}
The change of integration variables
\begin{equation}
x = \frac{\lambda}{2\sqrt{c-1}} \,,
\label{kpqr}
\end{equation}
leads to
\begin{equation}
a_0 =  \int_{-2 \sqrt{c-1}}^{2 \sqrt{c-1}} d \lambda \, f\left(\frac{\lambda}{\sqrt{c-1}}\right)  \rho_{0}(\lambda) \,,
\label{jpq}
\end{equation}
where $\rho_{0}(\lambda)$ is the leading contribution to the
eigenvalue distribution, as presented in eq. (\ref{KMfg}).
We do not need to compute explicitly the above integral, since
eq. (\ref{jpq}) is the Stieltjes transform of the eigenvalue
distribution $\rho_{0}(\lambda)$: this is nothing more than the definition
of the averaged resolvent. In order to evaluate the $O(1/N)$ correction, we need
to write down the coefficients $a_1,\dots,a_{\infty}$ in the same integral form.

By making the change of variables (\ref{kpqr}) in eqs. (\ref{re4}) and (\ref{re5}), we get
\begin{align}
a_{2j +1} &= \int_{-2 \sqrt{c-1}}^{2 \sqrt{c-1}} d \lambda \, f\left(\frac{\lambda}{\sqrt{c-1}}\right) C(\lambda) \,
T_{2 j +1}\left( \frac{\lambda}{2 \sqrt{c-1}}  \right) \,, \label{lk1} \\
a_{2j} &= \int_{-2 \sqrt{c-1}}^{2 \sqrt{c-1}} d \lambda \, f\left(\frac{\lambda}{\sqrt{c-1}}\right) C(\lambda) \, 
T_{2 j}\left( \frac{\lambda}{2 \sqrt{c-1}}  \right) \,, \label{lk2}
\end{align}
with $C(\lambda)$ defined by eq. (\ref{Cdef1}).
Now it is convenient to introduce, for $|\lambda| < 2 \sqrt{c-1}$, the function $g_c(\lambda)$
\begin{equation}
g_c(\lambda) = \frac{1}{2 (c-1)} \left( \lambda + i \sqrt{4 (c-1) - \lambda^2}     \right) \,, 
\end{equation}
which allows to rewrite, using eq. (\ref{laqp}), the Chebyshev polynomials as follows
\begin{equation}
T_{j}\left( \frac{\lambda}{2 \sqrt{c-1}}  \right) = (c-1)^{j/2} {\rm Re} \left[ g_c(\lambda)  \right]^{j} \,.
\end{equation}
Substituting this form of $T_j$ in eqs. (\ref{lk1}) and (\ref{lk2}), and then inserting the
resulting expressions in eq. (\ref{kl}), we obtain
\begin{align}
\frac{1}{N} \left\langle \sum_{i=1}^N f(\lambda_i) \right\rangle
&= \int_{-2 \sqrt{c-1}}^{2 \sqrt{c-1}} d \lambda \, f\left(\frac{\lambda}{\sqrt{c-1}}\right) \\
& \times
\left[ \rho_{0}(\lambda) + \frac{1}{N} \rho_1(\lambda)   \right] \,,
\label{klfe}
\end{align}
where
\begin{equation}
\rho_1(\lambda) = C(\lambda) {\rm Re} \left[\sum_{L=3}^{\infty}  V_L  g_c^L    \right] \,.
\label{lpqgu}
\end{equation}
The above summation starts at $L=3$, because $V_1$ and $V_2$
are zero. 

The $O(1/N)$ contribution in eq. (\ref{klfe}) is the Stieltjes transform 
of $\rho_1(\lambda)$, which yields the finite size correction $R_1(z)$ to the 
averaged resolvent. In order to compare with our results derived through the 
replica method, we rewrite eq. (\ref{rhofim}) according to
\begin{equation}
\rho_1(\lambda)  
= C(\lambda) {\rm Re} \left[\sum_{L=3}^{\infty}  (c-1)^L  \sum_{n=0}^{\infty} g_c^{L (n+1)}     \right]\,.
\label{lpqjdg}
\end{equation}
By comparing the coefficients $V_L$, defined by eq. (\ref{lpqde}), with those of the above equation, we
conclude that eqs. (\ref{lpqgu}) and (\ref{lpqjdg}) are the same.
  
\bibliography{bibliography.bib}

\begin{thebibliography}{39}
\expandafter\ifx\csname natexlab\endcsname\relax\def\natexlab#1{#1}\fi
\expandafter\ifx\csname bibnamefont\endcsname\relax
  \def\bibnamefont#1{#1}\fi
\expandafter\ifx\csname bibfnamefont\endcsname\relax
  \def\bibfnamefont#1{#1}\fi
\expandafter\ifx\csname citenamefont\endcsname\relax
  \def\citenamefont#1{#1}\fi
\expandafter\ifx\csname url\endcsname\relax
  \def\url#1{\texttt{#1}}\fi
\expandafter\ifx\csname urlprefix\endcsname\relax\def\urlprefix{URL }\fi
\providecommand{\bibinfo}[2]{#2}
\providecommand{\eprint}[2][]{\url{#2}}

\bibitem[{\citenamefont{D.~Cvetkovi\'c and Simi\'c}(2010)}]{Cvetbook}
\bibinfo{author}{\bibfnamefont{P.~R.} \bibnamefont{D.~Cvetkovi\'c}}
  \bibnamefont{and} \bibinfo{author}{\bibfnamefont{S.}~\bibnamefont{Simi\'c}},
  \emph{\bibinfo{title}{An introduction to the theory of graph spectra}}
  (\bibinfo{publisher}{Cambridge Univ. Press}, \bibinfo{address}{Cambridge},
  \bibinfo{year}{2010}).

\bibitem[{\citenamefont{Cavagna et~al.}(1999)\citenamefont{Cavagna, Giardina,
  and Parisi}}]{giorgio99}
\bibinfo{author}{\bibfnamefont{A.}~\bibnamefont{Cavagna}},
  \bibinfo{author}{\bibfnamefont{I.}~\bibnamefont{Giardina}}, \bibnamefont{and}
  \bibinfo{author}{\bibfnamefont{G.}~\bibnamefont{Parisi}},
  \bibinfo{journal}{Phys. Rev. Lett.} \textbf{\bibinfo{volume}{83}},
  \bibinfo{pages}{108} (\bibinfo{year}{1999}),
  \urlprefix\url{http://link.aps.org/doi/10.1103/PhysRevLett.83.108}.

\bibitem[{\citenamefont{Abou-Chacra et~al.}(1973)\citenamefont{Abou-Chacra,
  Thouless, and Anderson}}]{Abouchacra73}
\bibinfo{author}{\bibfnamefont{R.}~\bibnamefont{Abou-Chacra}},
  \bibinfo{author}{\bibfnamefont{D.~J.} \bibnamefont{Thouless}},
  \bibnamefont{and} \bibinfo{author}{\bibfnamefont{P.~W.}
  \bibnamefont{Anderson}}, \bibinfo{journal}{Journal of Physics C: Solid State
  Physics} \textbf{\bibinfo{volume}{6}}, \bibinfo{pages}{1734}
  (\bibinfo{year}{1973}),
  \urlprefix\url{http://stacks.iop.org/0022-3719/6/i=10/a=009}.

\bibitem[{\citenamefont{Baxter}(1982)}]{baxter1982}
\bibinfo{author}{\bibfnamefont{R.}~\bibnamefont{Baxter}},
  \emph{\bibinfo{title}{Exactly Solved Models in Statistical Mechanics}}
  (\bibinfo{publisher}{Academic Press}, \bibinfo{year}{1982}), ISBN
  \bibinfo{isbn}{9780120831821},
  \urlprefix\url{http://books.google.it/books?id=u\_JkQgAACAAJ}.

\bibitem[{\citenamefont{Rogers}(2010)}]{TimTese}
\bibinfo{author}{\bibfnamefont{T.}~\bibnamefont{Rogers}}, \bibinfo{journal}{PhD
  thesis, King's College, London}  (\bibinfo{year}{2010}).

\bibitem[{\citenamefont{Kabashima et~al.}(2010)\citenamefont{Kabashima,
  Takahashi, and Watanabe}}]{Kabashima2010}
\bibinfo{author}{\bibfnamefont{Y.}~\bibnamefont{Kabashima}},
  \bibinfo{author}{\bibfnamefont{H.}~\bibnamefont{Takahashi}},
  \bibnamefont{and} \bibinfo{author}{\bibfnamefont{O.}~\bibnamefont{Watanabe}},
  \bibinfo{journal}{Journal of Physics: Conference Series}
  \textbf{\bibinfo{volume}{233}}, \bibinfo{pages}{012001}
  (\bibinfo{year}{2010}),
  \urlprefix\url{http://stacks.iop.org/1742-6596/233/i=1/a=012001}.

\bibitem[{\citenamefont{Dumitriu and Pal}(2012)}]{Dumi2012}
\bibinfo{author}{\bibfnamefont{I.}~\bibnamefont{Dumitriu}} \bibnamefont{and}
  \bibinfo{author}{\bibfnamefont{S.}~\bibnamefont{Pal}}, \bibinfo{journal}{Ann.
  Probab.} \textbf{\bibinfo{volume}{40}}, \bibinfo{pages}{1861}
  (\bibinfo{year}{2012}).

\bibitem[{\citenamefont{Kesten}(1959)}]{Kesten}
\bibinfo{author}{\bibfnamefont{H.}~\bibnamefont{Kesten}},
  \bibinfo{journal}{Trans. Amer. Math. Soc.} \textbf{\bibinfo{volume}{92}},
  \bibinfo{pages}{336} (\bibinfo{year}{1959}).

\bibitem[{\citenamefont{McKay}(1981)}]{McKay}
\bibinfo{author}{\bibfnamefont{B.~D.} \bibnamefont{McKay}},
  \bibinfo{journal}{Linear Algebra Appl.} \textbf{\bibinfo{volume}{40}},
  \bibinfo{pages}{203} (\bibinfo{year}{1981}).

\bibitem[{\citenamefont{Efetov}(1990)}]{Efetov90}
\bibinfo{author}{\bibfnamefont{K.~B.} \bibnamefont{Efetov}},
  \bibinfo{journal}{Physica A} \textbf{\bibinfo{volume}{167}},
  \bibinfo{pages}{119} (\bibinfo{year}{1990}).

\bibitem[{\citenamefont{Montanari and Rizzo}(2005)}]{Montanari2005}
\bibinfo{author}{\bibfnamefont{A.}~\bibnamefont{Montanari}} \bibnamefont{and}
  \bibinfo{author}{\bibfnamefont{T.}~\bibnamefont{Rizzo}},
  \bibinfo{journal}{Journal of Statistical Mechanics: Theory and Experiment}
  \textbf{\bibinfo{volume}{2005}}, \bibinfo{pages}{P10011}
  (\bibinfo{year}{2005}),
  \urlprefix\url{http://stacks.iop.org/1742-5468/2005/i=10/a=P10011}.

\bibitem[{\citenamefont{Parisi and Slanina}(2006)}]{giorgio2006}
\bibinfo{author}{\bibfnamefont{G.}~\bibnamefont{Parisi}} \bibnamefont{and}
  \bibinfo{author}{\bibfnamefont{F.}~\bibnamefont{Slanina}},
  \bibinfo{journal}{Journal of Statistical Mechanics: Theory and Experiment}
  \textbf{\bibinfo{volume}{2006}}, \bibinfo{pages}{L02003}
  (\bibinfo{year}{2006}),
  \urlprefix\url{http://stacks.iop.org/1742-5468/2006/i=02/a=L02003}.

\bibitem[{\citenamefont{Sacksteder}(2007)}]{vincent2007}
\bibinfo{author}{\bibfnamefont{V.~E.} \bibnamefont{Sacksteder}},
  \bibinfo{journal}{Phys. Rev. D} \textbf{\bibinfo{volume}{76}},
  \bibinfo{pages}{105032} (\bibinfo{year}{2007}),
  \urlprefix\url{http://link.aps.org/doi/10.1103/PhysRevD.76.105032}.

\bibitem[{\citenamefont{Edwards and Jones}(1976)}]{Edwards76}
\bibinfo{author}{\bibfnamefont{S.~F.} \bibnamefont{Edwards}} \bibnamefont{and}
  \bibinfo{author}{\bibfnamefont{R.~C.} \bibnamefont{Jones}},
  \bibinfo{journal}{Journal of Physics A: Mathematical and General}
  \textbf{\bibinfo{volume}{9}}, \bibinfo{pages}{1595} (\bibinfo{year}{1976}),
  \urlprefix\url{http://stacks.iop.org/0305-4470/9/i=10/a=011}.

\bibitem[{\citenamefont{Rodgers and Bray}(1988)}]{RB88}
\bibinfo{author}{\bibfnamefont{G.~J.} \bibnamefont{Rodgers}} \bibnamefont{and}
  \bibinfo{author}{\bibfnamefont{A.~J.} \bibnamefont{Bray}},
  \bibinfo{journal}{Phys. Rev. B} \textbf{\bibinfo{volume}{37}},
  \bibinfo{pages}{3557} (\bibinfo{year}{1988}),
  \urlprefix\url{http://link.aps.org/doi/10.1103/PhysRevB.37.3557}.

\bibitem[{\citenamefont{Ferrari et~al.}(2013)\citenamefont{Ferrari, Lucibello,
  Morone, Parisi, Ricci-Tersenghi, and Rizzo}}]{Carlo2013}
\bibinfo{author}{\bibfnamefont{U.}~\bibnamefont{Ferrari}},
  \bibinfo{author}{\bibfnamefont{C.}~\bibnamefont{Lucibello}},
  \bibinfo{author}{\bibfnamefont{F.}~\bibnamefont{Morone}},
  \bibinfo{author}{\bibfnamefont{G.}~\bibnamefont{Parisi}},
  \bibinfo{author}{\bibfnamefont{F.}~\bibnamefont{Ricci-Tersenghi}},
  \bibnamefont{and} \bibinfo{author}{\bibfnamefont{T.}~\bibnamefont{Rizzo}},
  \bibinfo{journal}{Phys. Rev. B} \textbf{\bibinfo{volume}{88}},
  \bibinfo{pages}{184201} (\bibinfo{year}{2013}),
  \urlprefix\url{http://link.aps.org/doi/10.1103/PhysRevB.88.184201}.

\bibitem[{\citenamefont{Lucibello et~al.}(2014)\citenamefont{Lucibello, Morone,
  Parisi, Ricci-Tersenghi, and Rizzo}}]{Carlo2014}
\bibinfo{author}{\bibfnamefont{C.}~\bibnamefont{Lucibello}},
  \bibinfo{author}{\bibfnamefont{F.}~\bibnamefont{Morone}},
  \bibinfo{author}{\bibfnamefont{G.}~\bibnamefont{Parisi}},
  \bibinfo{author}{\bibfnamefont{F.}~\bibnamefont{Ricci-Tersenghi}},
  \bibnamefont{and} \bibinfo{author}{\bibfnamefont{T.}~\bibnamefont{Rizzo}},
  \bibinfo{journal}{Phys. Rev. E} \textbf{\bibinfo{volume}{90}},
  \bibinfo{pages}{012146} (\bibinfo{year}{2014}),
  \urlprefix\url{http://link.aps.org/doi/10.1103/PhysRevE.90.012146}.

\bibitem[{\citenamefont{Arous and Dang}(2011)}]{BenArous2011}
\bibinfo{author}{\bibfnamefont{G.~B.} \bibnamefont{Arous}} \bibnamefont{and}
  \bibinfo{author}{\bibfnamefont{K.}~\bibnamefont{Dang}},
  \bibinfo{journal}{arXiv:1106.2108}  (\bibinfo{year}{2011}).

\bibitem[{\citenamefont{Dumitriu et~al.}(2013)\citenamefont{Dumitriu, Johnson,
  Pal, and Paquette}}]{Dumitriu2013}
\bibinfo{author}{\bibfnamefont{I.}~\bibnamefont{Dumitriu}},
  \bibinfo{author}{\bibfnamefont{T.}~\bibnamefont{Johnson}},
  \bibinfo{author}{\bibfnamefont{S.}~\bibnamefont{Pal}}, \bibnamefont{and}
  \bibinfo{author}{\bibfnamefont{E.}~\bibnamefont{Paquette}},
  \bibinfo{journal}{Probability Theory and Related Fields}
  \textbf{\bibinfo{volume}{156}}, \bibinfo{pages}{921} (\bibinfo{year}{2013}),
  ISSN \bibinfo{issn}{0178-8051},
  \urlprefix\url{http://dx.doi.org/10.1007/s00440-012-0447-y}.

\bibitem[{\citenamefont{Johnson}(2012)}]{Johnson2012}
\bibinfo{author}{\bibfnamefont{T.}~\bibnamefont{Johnson}},
  \bibinfo{journal}{arXiv:1112.0704}  (\bibinfo{year}{2012}).

\bibitem[{\citenamefont{Dean}(2002)}]{Dean}
\bibinfo{author}{\bibfnamefont{D.~S.} \bibnamefont{Dean}},
  \bibinfo{journal}{Journal of Physics A: Mathematical and General}
  \textbf{\bibinfo{volume}{35}}, \bibinfo{pages}{L153} (\bibinfo{year}{2002}),
  \urlprefix\url{http://stacks.iop.org/0305-4470/35/i=12/a=101}.

\bibitem[{\citenamefont{K\"uhn}(2008)}]{K08}
\bibinfo{author}{\bibfnamefont{R.}~\bibnamefont{K\"uhn}}, \bibinfo{journal}{J.
  Phys. A: Math. Theor.} \textbf{\bibinfo{volume}{41}}, \bibinfo{pages}{295002}
  (\bibinfo{year}{2008}).

\bibitem[{\citenamefont{Erg\"un and K\"uhn}(2009)}]{Kuhn2009}
\bibinfo{author}{\bibfnamefont{G.}~\bibnamefont{Erg\"un}} \bibnamefont{and}
  \bibinfo{author}{\bibfnamefont{R.}~\bibnamefont{K\"uhn}},
  \bibinfo{journal}{Journal of Physics A: Mathematical and Theoretical}
  \textbf{\bibinfo{volume}{42}}, \bibinfo{pages}{395001}
  (\bibinfo{year}{2009}),
  \urlprefix\url{http://stacks.iop.org/1751-8121/42/i=39/a=395001}.

\bibitem[{\citenamefont{Rogers et~al.}(2010)\citenamefont{Rogers, Vicente,
  Takeda, and Castillo}}]{Tim2010}
\bibinfo{author}{\bibfnamefont{T.}~\bibnamefont{Rogers}},
  \bibinfo{author}{\bibfnamefont{C.~P.} \bibnamefont{Vicente}},
  \bibinfo{author}{\bibfnamefont{K.}~\bibnamefont{Takeda}}, \bibnamefont{and}
  \bibinfo{author}{\bibfnamefont{I.~P.} \bibnamefont{Castillo}},
  \bibinfo{journal}{Journal of Physics A: Mathematical and Theoretical}
  \textbf{\bibinfo{volume}{43}}, \bibinfo{pages}{195002}
  (\bibinfo{year}{2010}),
  \urlprefix\url{http://stacks.iop.org/1751-8121/43/i=19/a=195002}.

\bibitem[{\citenamefont{K\"uhn and van Mourik}(2011)}]{Kuhn2011}
\bibinfo{author}{\bibfnamefont{R.}~\bibnamefont{K\"uhn}} \bibnamefont{and}
  \bibinfo{author}{\bibfnamefont{J.}~\bibnamefont{van Mourik}},
  \bibinfo{journal}{Journal of Physics A: Mathematical and Theoretical}
  \textbf{\bibinfo{volume}{44}}, \bibinfo{pages}{165205}
  (\bibinfo{year}{2011}),
  \urlprefix\url{http://stacks.iop.org/1751-8121/44/i=16/a=165205}.

\bibitem[{\citenamefont{Nagao}(2013)}]{Nagao2013}
\bibinfo{author}{\bibfnamefont{T.}~\bibnamefont{Nagao}},
  \bibinfo{journal}{Journal of Physics A: Mathematical and Theoretical}
  \textbf{\bibinfo{volume}{46}}, \bibinfo{pages}{065003}
  (\bibinfo{year}{2013}),
  \urlprefix\url{http://stacks.iop.org/1751-8121/46/i=6/a=065003}.

\bibitem[{\citenamefont{Bordenave and Lelarge}(2010)}]{bordenave}
\bibinfo{author}{\bibfnamefont{C.}~\bibnamefont{Bordenave}} \bibnamefont{and}
  \bibinfo{author}{\bibfnamefont{M.}~\bibnamefont{Lelarge}},
  \bibinfo{journal}{Random Structures \& Algorithms}
  \textbf{\bibinfo{volume}{37}}, \bibinfo{pages}{332} (\bibinfo{year}{2010}),
  ISSN \bibinfo{issn}{1098-2418},
  \urlprefix\url{http://dx.doi.org/10.1002/rsa.20313}.

\bibitem[{\citenamefont{Metz et~al.}(2010)\citenamefont{Metz, Neri, and
  Boll\'e}}]{Metz2010}
\bibinfo{author}{\bibfnamefont{F.~L.} \bibnamefont{Metz}},
  \bibinfo{author}{\bibfnamefont{I.}~\bibnamefont{Neri}}, \bibnamefont{and}
  \bibinfo{author}{\bibfnamefont{D.}~\bibnamefont{Boll\'e}},
  \bibinfo{journal}{Phys. Rev. E} \textbf{\bibinfo{volume}{82}},
  \bibinfo{pages}{031135} (\bibinfo{year}{2010}),
  \urlprefix\url{http://link.aps.org/doi/10.1103/PhysRevE.82.031135}.

\bibitem[{\citenamefont{Economou}(2006)}]{Economoubook1}
\bibinfo{author}{\bibfnamefont{E.~N.} \bibnamefont{Economou}},
  \emph{\bibinfo{title}{Green's functions in quantum physics}}
  (\bibinfo{publisher}{Springer}, \bibinfo{address}{Heidelberg},
  \bibinfo{year}{2006}).

\bibitem[{\citenamefont{Bollob\'as}(1980)}]{Bollobas80}
\bibinfo{author}{\bibfnamefont{B.}~\bibnamefont{Bollob\'as}},
  \bibinfo{journal}{European Journal of Combinatorics}
  \textbf{\bibinfo{volume}{1}}, \bibinfo{pages}{311} (\bibinfo{year}{1980}).

\bibitem[{\citenamefont{Wormald}(1981)}]{Wormald81}
\bibinfo{author}{\bibfnamefont{N.~C.} \bibnamefont{Wormald}},
  \bibinfo{journal}{Journal of Combinatorial Theory Series B}
  \textbf{\bibinfo{volume}{31}}, \bibinfo{pages}{168} (\bibinfo{year}{1981}).

\bibitem[{\citenamefont{Davis and Rabinowitz}(1984)}]{Philip}
\bibinfo{author}{\bibfnamefont{P.~J.} \bibnamefont{Davis}} \bibnamefont{and}
  \bibinfo{author}{\bibfnamefont{P.}~\bibnamefont{Rabinowitz}},
  \emph{\bibinfo{title}{Methods of Numerical Integration}}
  (\bibinfo{publisher}{Academic Press}, \bibinfo{address}{London},
  \bibinfo{year}{1984}).

\bibitem[{\citenamefont{Waldvogel}(2011)}]{Waldvogel2011}
\bibinfo{author}{\bibfnamefont{J.}~\bibnamefont{Waldvogel}},
  \emph{\bibinfo{title}{Towards a General Error Theory of the Trapezoidal Rule
  in Approximation and Computation}}, vol.~\bibinfo{volume}{42}
  (\bibinfo{publisher}{Springer New York}, \bibinfo{year}{2011}).

\bibitem[{\citenamefont{Trefethen and Weideman}(2014)}]{Loyd}
\bibinfo{author}{\bibfnamefont{L.~N.} \bibnamefont{Trefethen}}
  \bibnamefont{and} \bibinfo{author}{\bibfnamefont{J.~A.~C.}
  \bibnamefont{Weideman}}, \bibinfo{journal}{SIAM Review}
  \textbf{\bibinfo{volume}{56}}, \bibinfo{pages}{385} (\bibinfo{year}{2014}).

\bibitem[{\citenamefont{Hardy}(1991)}]{hardy1991}
\bibinfo{author}{\bibfnamefont{G.}~\bibnamefont{Hardy}},
  \emph{\bibinfo{title}{Divergent Series}}, Chelsea Publishing Series
  (\bibinfo{publisher}{American Mathematical Society}, \bibinfo{year}{1991}),
  ISBN \bibinfo{isbn}{9780821826492},
  \urlprefix\url{http://books.google.it/books?id=jPccoUKsLdQC}.

\bibitem[{\citenamefont{Steger and Wormald}(1999)}]{WormaldA}
\bibinfo{author}{\bibfnamefont{A.}~\bibnamefont{Steger}} \bibnamefont{and}
  \bibinfo{author}{\bibfnamefont{N.~C.} \bibnamefont{Wormald}},
  \bibinfo{journal}{Comb. Probab. Comput.} \textbf{\bibinfo{volume}{8}},
  \bibinfo{pages}{377} (\bibinfo{year}{1999}), ISSN \bibinfo{issn}{0963-5483},
  \urlprefix\url{http://dx.doi.org/10.1017/S0963548399003867}.

\bibitem[{\citenamefont{Verbaarschot and Zirnbauer}(1984)}]{Verb84}
\bibinfo{author}{\bibfnamefont{J.~J.~M.} \bibnamefont{Verbaarschot}}
  \bibnamefont{and} \bibinfo{author}{\bibfnamefont{M.~R.}
  \bibnamefont{Zirnbauer}}, \bibinfo{journal}{Ann.Phys.}
  \textbf{\bibinfo{volume}{158}}, \bibinfo{pages}{78} (\bibinfo{year}{1984}).

\bibitem[{\citenamefont{Dhesi and Jones}(1990)}]{Dhesi90}
\bibinfo{author}{\bibfnamefont{G.~S.} \bibnamefont{Dhesi}} \bibnamefont{and}
  \bibinfo{author}{\bibfnamefont{R.~C.} \bibnamefont{Jones}},
  \bibinfo{journal}{Journal of Physics A: Mathematical and General}
  \textbf{\bibinfo{volume}{23}}, \bibinfo{pages}{5577} (\bibinfo{year}{1990}),
  \urlprefix\url{http://stacks.iop.org/0305-4470/23/i=23/a=029}.

\bibitem[{\citenamefont{Mirlin and Fyodorov}(1991)}]{fyodorov91}
\bibinfo{author}{\bibfnamefont{A.~D.} \bibnamefont{Mirlin}} \bibnamefont{and}
  \bibinfo{author}{\bibfnamefont{Y.~V.} \bibnamefont{Fyodorov}},
  \bibinfo{journal}{Journal of Physics A: Mathematical and General}
  \textbf{\bibinfo{volume}{24}}, \bibinfo{pages}{2273} (\bibinfo{year}{1991}),
  \urlprefix\url{http://stacks.iop.org/0305-4470/24/i=10/a=016}.

\end{thebibliography}

\end{document}